\documentclass[%
reprint,
 amsmath,amssymb,
 aps
]{revtex4-2}

\usepackage{graphicx}
\usepackage{dcolumn}
\usepackage{soul}
\usepackage{bm}

\usepackage{amsmath}
\usepackage{graphicx}
\usepackage{amsfonts}
\usepackage{color}
\usepackage{soul}
\usepackage{wrapfig}
\usepackage{titlecaps}
\usepackage{ragged2e}
\usepackage{pifont}
\usepackage{amsmath}
\usepackage{nccmath}
\usepackage{graphicx}
\graphicspath{ C:\Users\jdubi\Documents\PhD\Project ManyBodySystems\Paper\Draft1}
\usepackage{epstopdf}
\usepackage{float}
\usepackage{wrapfig}
\usepackage{setspace}
\usepackage{geometry}
\usepackage{braket}
\usepackage{lineno}
\usepackage{xfrac}

\makeatletter
\def\captionof#1#2{{\def\@captype{#1}#2}}
\makeatother
\def\beq{\begin{equation}}
\def\eeq{\end{equation}}
\def\beqa{\begin{eqnarray}}
\def\eeqa{\end{eqnarray}}

\newcommand{\appropto}{\mathrel{\vcenter{
			\offinterlineskip\halign{\hfil$##$\cr
				\propto\cr\noalign{\kern2pt}\sim\cr\noalign{\kern-2pt}}}}}
\newcommand{\prlsec}[1]{{\sl #1}.--}

\begin{document}


\title{Effects of Disorder and Interactions in Environment Assisted Quantum Transport}
\author{Elinor Zerah-Harush}
 \affiliation{Department of Chemistry and the Ilse Katz Center for nanoscale Science and Technology, Ben-Gurion University of the Negev, Beer Sheva, Israel}
\author{Yonatan Dubi}
 \email{jdubi@bgu.ac.il}
\affiliation{Department of Chemistry and the Ilse Katz Center for nanoscale Science and Technology, Ben-Gurion University of the Negev, Beer Sheva, Israel}

\begin{abstract}
Understanding the interplay between disorder, environment and interactions is key to elucidating the transport properties of open quantum systems, from excitons in photosynthetic networks to qubits in ion traps. This interplay is studied here theoretically in the context of environment-assisted quantum transport (ENAQT), a unique situation in open system where an environment-induced dephasing can, counter-intuitively, enhance transport. First, we show a surprising situation where 
the particle current grows with increasing disorder, even without dephasing. Then, we suggest a specific mechanism for ENAQT (which we dub population uniformization) and demonstrate that it can explain the persistence of ENAQT deep into the disorder-induced localization regime. Finally, we show that repulsive interactions are detrimental to ENAQT, and lead to an environment-hampered quantum transport. Our predictions can readily be tested within the scope of particle current experimental capabilities. 
\end{abstract}

\maketitle


\prlsec{Introduction} 
Environment assisted quantum transport (ENAQT) is a phenomenon that occurs when transport of quantum particles (excitons, spins, electrons etc.) is interrupted by an environment in such a way that, counter intuitively, transfer efficiency is enhanced.  ENAQT was proposed as the origin of the exceptional performances of energy transport in photosynthetic systems \cite{caruso2009,panitchayangkoon2010,kassal2013does}. Since then, ENAQT was studied extensively \cite{levi2015,caruso2009,Chin2012,Kassal2012,Ishizaki2012,Fleming2011,Cheng2009,Dawlaty2012,Collini2013,Lambert2013,Scholes2005,Pachon2012,Scholes2014}, and observed in many other systems such as classical oscillator networks \cite{Aragón2015}, photonic latices \cite{viciani2016, caruso2016}, and very recently in qubit chains \cite{maier2019}. These latter experiments present unprecedented control over transport and system-environment coupling in open quantum systems, and demonstrate the hallmark of ENAQT, namely a non-monotonic dependence of particle transport efficiency on the environment-induced dephasing rate. 
\\

\begin{figure}
    \centering
    \includegraphics[width=7.5truecm]{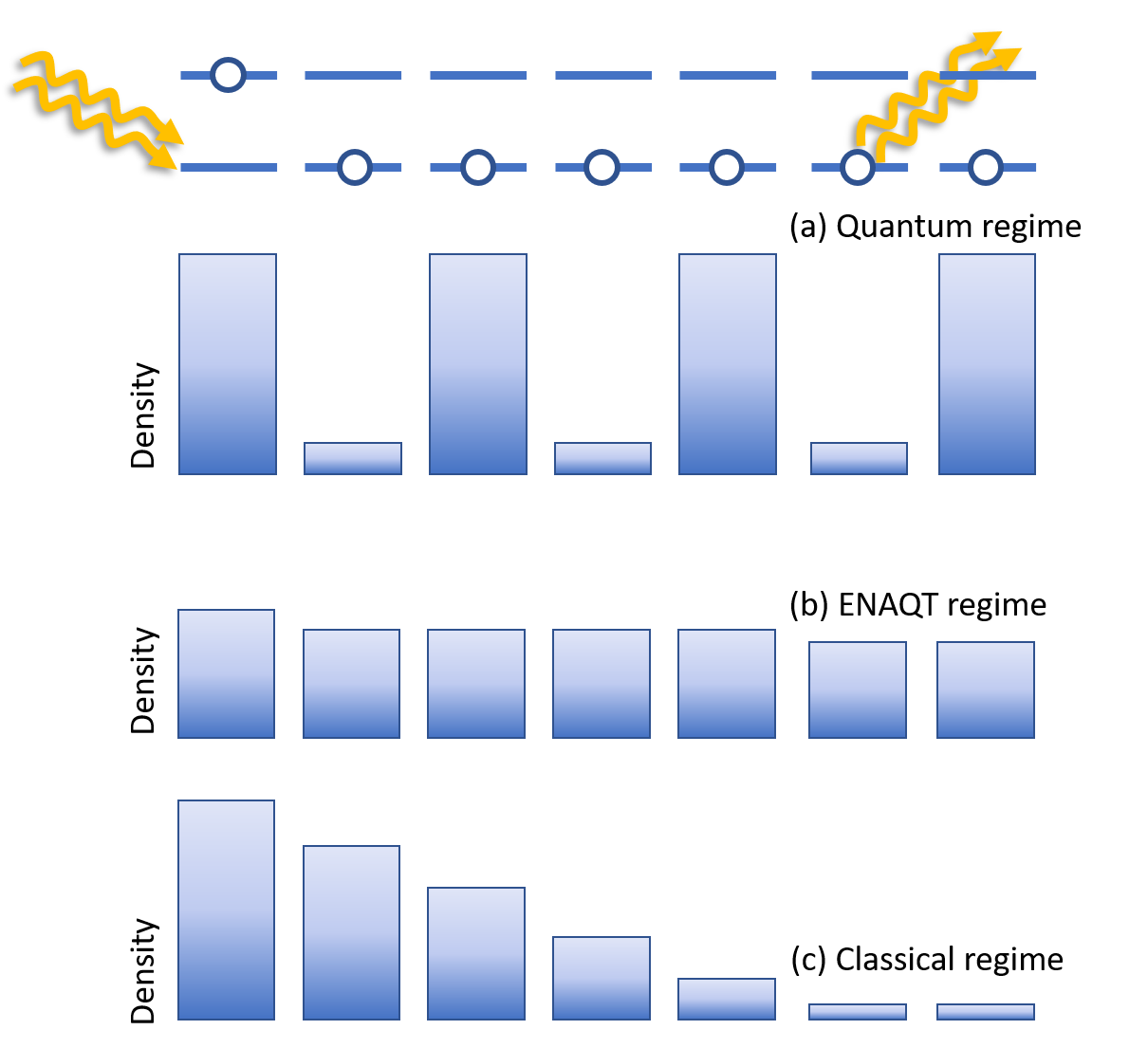}
    \caption{
    schematic presentation of the population uniformization mechanism for ENAQT \cite{zerah2018}. In the top image, Blue bars represent energy levels of a seven-site chain coupled to a source (left-most) and drain (6th) sites. A particle (e.g. exciton) is excited in the source site (e.g. via photon absorption) and is released to the drain site (e.g. via photon emission). Blue bars describe particle population along the chain, under different dephasing conditions. (a)Quantum regime: population distribution is dominated by the Hamiltonian wave functions and the locations of source and drain. (b) ENAQT regime: Introducing dephasing to the system results in uniformization of the population that increases the population at the extraction site, and hence raises the particle current at this point. (c) Classical regime: at high dephasing rates, the system goes into the classic limit, characterized by Fick's law and a generation of a density gradient.       }
    \label{fig.schematic} 
\end{figure}

Recently, we showed that ENAQT is due to a mechanism we here dub "population uniformization". \cite{zerah2018} To understand this mechanism, schematically depicted in Fig.~ \ref{fig.schematic}, consider a chain connected to a source and drain (as indeed was manifested experimentally in Ref.~\cite{maier2019}), we note that the particle current (and the transport efficiency) are proportional to the particle density at the drain sites\cite{zerah2018}. When the system is coupled only to a particle source and drain (i.e. no dephasing due to an environment along the chain), transport is dominated by the Hamiltonian wave functions that are constrained by the source and drain locations (Fig.~ \ref{fig.schematic}(a)). Adding an environment leads to two competing effects:  (1) A "Zeno effect"\cite{misra1977} where the local particle population is constantly measured, resulting in dephasing. consequently, the particle population (distributed throughout the system) becomes more uniform \cite{zerah2018} (Fig.~ \ref{fig.schematic}(b)). This results in an increase in the particle population at the drain site and the particle current increases. (2) Formation of a population gradient (Fig.~ \ref{fig.schematic}(c)), which is a manifestation of Fick's law at the onset of dephasing-induced classical dynamics.\cite{Meixner1965,Kubo1966} This gradient forces the density close to the sink to be reduced, which results in a reduction of particle current. The competition between the tendency of the dephasing environment to reduce the population fluctuations and the  generation of a population gradient results in the ENAQT phenomenon.

 Here, we report a study on the effects of disorder and interactions on ENAQT  in light of the above mechanism. We show that, surprisingly, weak disorder may assist the transport for low dephasing rates under some geometrical conditions. We then show that ENAQT is robust under increasing disorder, and occurs also for extreme case of strong disorder. Finally, we study the influence of interactions on ENAQT, and show that they may lead to a qualitative change in the transport properties, from ENAQT to environment-induced transport {\sl 
 reduction}.
With the advanced  experimental ability to control all the parameters in the system \cite{maier2019}, our predictions can be tested and verified experimentally. 
\\

\prlsec{Model and Method} The system is described by a general tight binding Hamiltonian,
\begin{equation}
H=\sum_i\epsilon_i e^\dagger_{i}e_{i}-\sum_{i,j}t_{i,j}e^\dagger_{i}e_{j}~~,
\end{equation}
where $i$ is a real-space position index and $e^\dagger_{i} (e_i)$ creates a particle (e.g. excitons in chromophores, or excited states in ions) in position $r_i$).  $\epsilon_i$ are the on-site energies, with $\epsilon_i=\epsilon_0+\xi_i$. $\xi_i$ are random energies which are drawn from a uniform distribution, $\xi_i\in U [-\frac{W}{2},\frac{W}{2} ]$, such that $W$ represents the disorder strength. $t_{ij}$ are hopping matrix elements, which are taken hereon to be nearest neighbors, $t_{ij}=t~  \delta_{i+1,j}$ (long range hopping, as in Ref.\cite{maier2019}, is discussed in appendix D).  

The density matrix $\rho$ is calculated though the Lindblad equation \cite{breuer2002theory}, 
\begin{eqnarray}\label{LindbladEq}
   \frac{ d\rho}{dt}&=&-i[H,\rho]+L[\rho]~, \nonumber\\
   L[\rho]&=&\sum_k \left(V_k\rho V^{\dagger}_k-\frac{1}{2} \left\{ V^{\dagger}_kV_k,\rho \right\}\right)~
\end{eqnarray}
where $V_k$ are a list of Lindblad operators encoding the action of the environment on the system and $\{\cdot,\cdot\}$ is the anti-commutation. The Lindblad operators for dephasing operate on each site via $V_i=(\gamma_{deph})^{1/2} e_i^{\dagger}e_i$. Exciton injection and extraction are defined by Lindblad operators $V_{inj}=(\gamma_{inj})^{1/2} e_{i_{inj}}^{\dagger} $ and $V_{ext}=(\gamma_{ext})^{1/2} e_{i_{ext}} $. $\gamma_{deph},~\gamma_{inj}$ and $\gamma_{ext}$ are the particle dephasing, injection and extraction rates, respectively, and $i_{inj}$ and $i_{ext}$ are the indices of the sites where injection and extraction take place, respectively. The steady-state density matrix $\rho_{S}$, from which all the relevant quantities (currents, density) , is evaluated as the kernel of the Lindbladian super-operator \cite{dubi2015}.

Typically, studies of  the ENAQT phenomenon 
in open quantum networks  treat the dynamics of the system (e.g.,Refs.~\cite{ritschel2011,Caruso2010,caruso2016,viciani2016,plenio2008}). Here, we choose to consider the steady-state properties of the system \cite{dubi2015,Manzano2013} (i.e. solving $\frac{d\rho}{dt}=0$ in Eq.~\ref{LindbladEq} to obtain $\rho_S$), in some similarity to studies of non-equilibrium steady states in boundary-driven open systems \cite{karevski2009quantum,karevski2013exact,prosen2015matrix}. This choice is based on several reasons; (i) The steady-state particle currents and populations give a clear demonstration of the "population uniformization" mechanism \cite{zerah2018}, which is hard to see from the dynamics, yet (ii) there is correspondence between the currents in the dynamic and steady-state cases \cite{dubi2015} (especially if the particle injection rate is small), and (iii) it seems that natural systems (e.g. photosynthetic networks) operate at the steady state with incoherent excitations. \cite{brumer2018shedding,axelrod2018efficient,Manzano2013}

The steady-state solution $\rho_{SS}$ is obtained by numerically evaluating the Kernel of the Lindbladian. In order to do this numerically efficiently, we restrict the Fock-space up to one exciton (two excitons for interacting excitons). 
Within the parameters we use (which are extracted from experiment), the exciton number $n_e$ (which is primarily related to the ratio between injection and extraction rates, see appendix A) is larger than one, thus one would argue that the full Fock space is needed. However, as we show in appendix B, the behavior we present here is similar to that of a system with $n_e<<1$ (where the single-exciton Fock state approximation is valid) with only quantitative differences, as indeed one would expect for non-interacting excitons.

Once the stead-state solution is obtained, the exciton current can be evaluated by \cite{dubi2015} 
\begin{equation}\label{eq:3001}
J=\mathrm{Tr}(\hat{N} L_{ext}[\rho_{SS}]).
\end{equation}
Where $\hat{N}=\sum_{i}e^\dagger_i e_i$ is the occupation matrix, and $L_{ext}[\rho_{SS}]$ is the extraction part of the Lindbladian (i.e. Eq.~\ref{LindbladEq} with $V_{ext}$ only). As shown in Ref.~\onlinecite{zerah2018}, within the Lindblad approximation the current is simply related to the density of excitons at the extraction sites, $J=\Gamma_{ext}\sum_{i_{ext}} \rho_{SS}(i_{ext},i_{ext})$. \\

\prlsec{Disorder without dephasing}
We start by discussing a unique situation which happens in open quantum chains that exhibit insignificant rate of dephasing, i.e. fully quantum chains. Note that they are still "open quantum systems", because they are coupled to a source and a drain. Common wisdom implies that adding (static) disorder tends to reduce the current through a quantum system, due to the onset of localization of the wave-functions. However, under certain conditions, weak disorder can enhance the current, and a maximum current can be obtained at a finite disorder strength (a similar effect was reported in Ref.~\cite{Rebentrost2009} discussing binary-tree-like networks, but its origin was not discussed). Adding an environment-induced dephasing diminished this effect, until it vanishes for high enough dephasing rate.   

\begin{figure}[H]
    \centering
    \includegraphics[width=8
truecm]{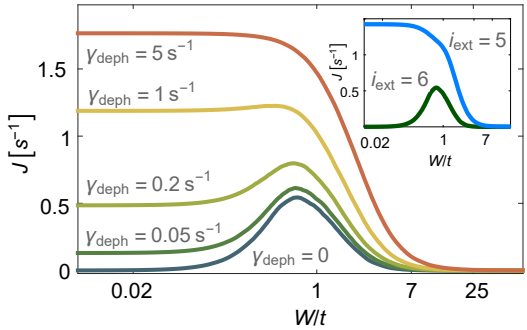}
    \caption{Particle current as a function of disorder for a system of seven sites (as in Fig.~\ref{fig.schematic}) for low rates of dephasing: 
    $\gamma_{deph}=0,0.05,0.2,1,5 s^{-1}$. Inset:  particle current as a function of disorder for a system that is coupled to the drain at site 6 (green line) and site 5 (blue line). Only when the position of the drain site is commensurate with a node in a wave-function the disorder-enhanced transport effect comes into play (see appendix C). Other parameters are: $\gamma_{inj}=\gamma_{ext}=17s^{-1},\epsilon=43000s^{-1},t=145s^{-1} $
    \label{fig:diff_deph} }
\end{figure}

To see this, in Fig.~ \ref{fig:diff_deph} we plot the particle current as a function of disorder (in units of the hopping matrix element), for different values of dephasing rate, $\gamma_{deph}=0,0.05,0.2,1,5 s^{-1}$ (green to red), averaged over $5000$ realizations of disorder. We set $\epsilon_0=43000 s^{-1}, t=145 s^{-1}, \gamma_{inj} =17s^{-1}$ and $\gamma_{ext}=17s^{-1}$. Parameters were chosen to fit the experiments of Ref.~\cite{maier2019}. The chain is of length $L=7$ and the extraction site is at $i_{ext}=6$ (and $i_{ext}=5$ in the inset). 

As can be seen, at zero dephasing (dark green line in Fig.~\ref{fig:diff_deph}), the particle current rises exponentially, having a maximum at $W/t\sim 1$, and then reduces back to its original value. As dephasing is added (green to red lines), the maximum becomes less pronounced, until it vanishes altogether for $\gamma_{deph}\approx 5s^{-1} $.

Remembering that particle current is proportional to the particle population at the extraction site, the origin of this disorder-enhanced transport can understood as follows (more detailed explanation appendix C). The steady state solution tends to minimize the particle population at the drain site. Thus, if one of the wave-functions of the system has a node in the drain site, that wave function will be the primary state occupied, i.e the system is localized in eigen-space. Introducing disorder to the Hamiltonian forces the system to occupy more eigen-states. Consequently, the exit site becomes more populated, hence particle current increases. Additional disorder to the system brings to decrease in overlap from source to drain, and as an outcome, the real-space localization of the wave-functions reduces the particle current.

On the other hand, if there is no node at the extraction site (such as blue line in the inset of Fig.~ \ref{fig:diff_deph}), the ordered system is delocalized in eigen-space, hence particle current is finite and disorder decreases it via real-space localization. When dephasing is added, it induces mixing between the eigen-states. Thus, if the system was localized in eigen-space in the absence of dephasing (i.e. if the extraction site overlays a node in one of the eigen-functions), it undergoes eigen-space delocalization, yielding finite current even for small disorder. Then, upon increasing the disorder strength, particle current is reduced via real-space localization. 

\begin{figure}[H]
    \centering
    \includegraphics[width=8truecm]{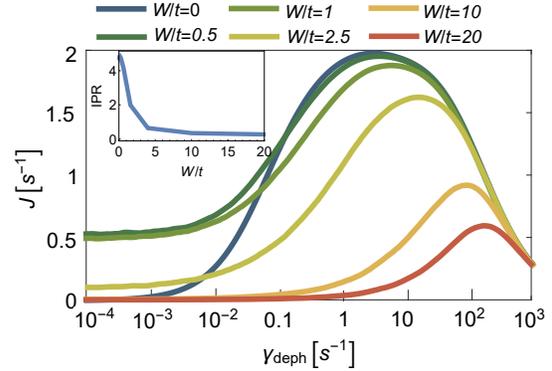}
    \caption{ENAQT in disordered systems: particle current as a function of dephasing rate for different disorder strengths $W/t= 0,0.5,1,2.5,10,20$. ENAQT appears even for very high disorder. Inset: averaged Inverse participation ratio as a function of disorder.}
    \label{fig:Disordered_ENAQT}
\end{figure}

\prlsec{Dephasing and Disorder: the robustness of ENAQT}
We now move to show the interplay between dephasing and disorder in generating ENQAT. 
Fig.~ \ref{fig:Disordered_ENAQT}  shows the particle current as a function of dephasing rate for different disorder strengths, $W/t=0,0.5,1,2.5,10,20$. As can be seen, the occurrence of ENAQT does not depend on the strength of disorder, as it appears for both weakly and highly disordered system (up to $W/t=20$ and beyond). The inset of Fig.~ \ref{fig:Disordered_ENAQT} shows the inverse participation ratio, $\mathrm{IPR}=\left( \sum_{n,i}|\psi_n(i)|^4 \right)^{-1} $, as a function of disorder, indicating that ENAQT persists even when the system is deep within the localized regime. This shows that the appearance of ENAQT is {\sl not a result of localization}, but can be grasped more accurately by the population uniformization mechanism.

\begin{widetext}
\begin{figure}[H]
    \centering
    \includegraphics[width=14.5truecm]{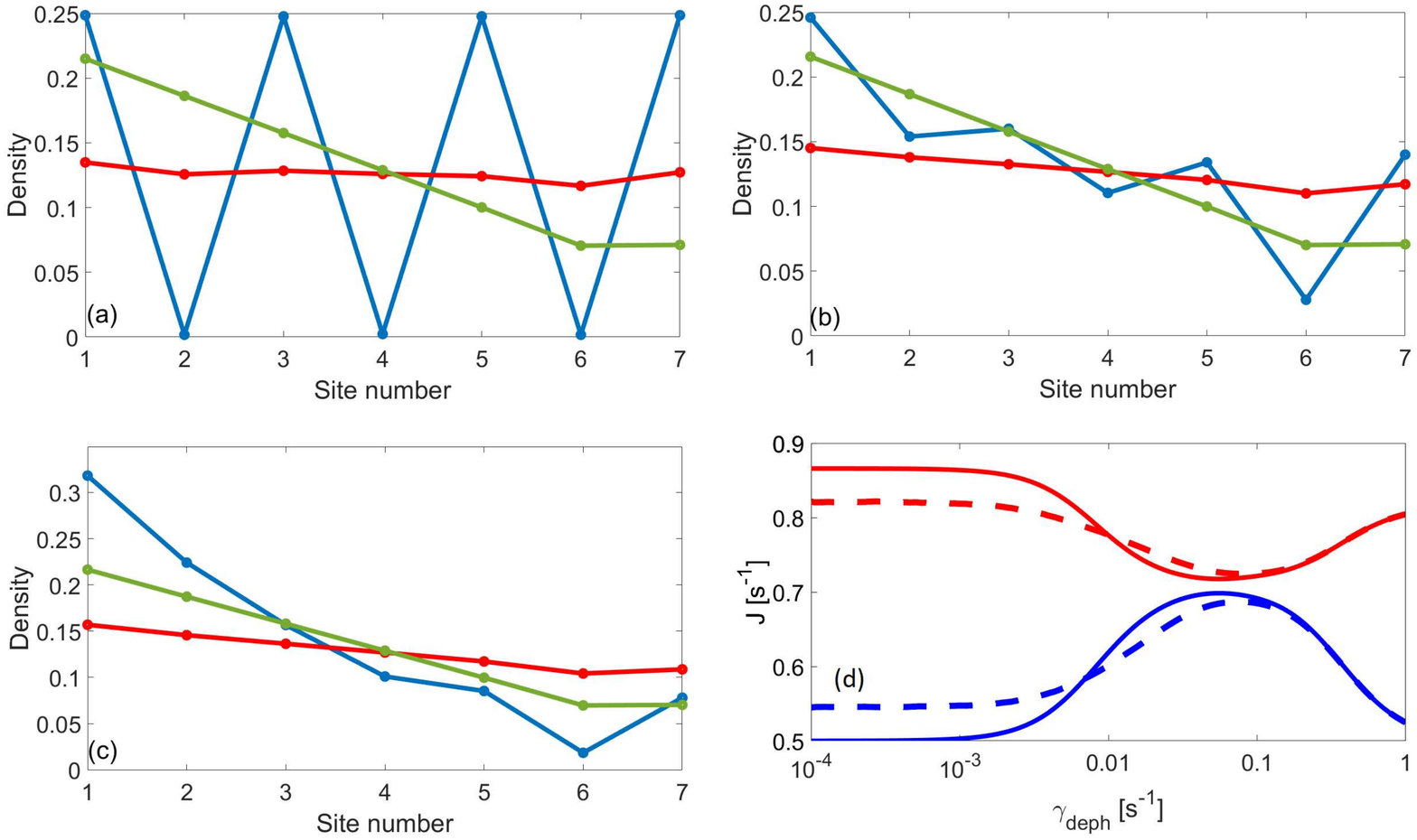}
    \caption{Influence of disorder on  ENAQT: Density (population) as a function of site location for different dephasing rates; low ($\gamma_{deph}\approx0$, that corresponds to quantum regime,blue), 
    ENAQT regime ($\gamma_{deph}=30,55,98 s^{-1}$ for a,b and c respectively, red) and high ($\gamma_{deph}=1000s^{-1}$, corresponds to classical regime, green) and for different strength of disorder, $W/t=0,1,4$ (a-c respectively). In the ordered system (a) the population uniformization mechanism is clear, going from a wave-like density distribution (quantum regime, blue) through a uniform distribution (ENAQT regime, red) to the classical Fick's law (green), in similarity to the schematic of Fig.~\ref{fig.schematic}. As disorder increases (b and c) the wave-structure of the population in the quantum regime is distorted, generating a population gradient driven by the wave-function localization and the positions of the source and sink. (d) Particle current (blue) and population spread parameter $\Delta_n$ (red line) for ordered (solid line) and disordered (dashed line) chains. Injection and extraction rates are as in figure 3. }
    \label{fig:ENAQT_Mechanism} 
\end{figure}
\end{widetext}

The population uniformization mechanism of ENAQT in disordered systems is demonstrated in Fig.~ \ref{fig:ENAQT_Mechanism}, where the particle population is shown as a function of sites along the chain for different values of disorder, $W/t=0,1,4$ (Figs.~\ref{fig:ENAQT_Mechanism} a,b and c, respectively) and for zero dephasing (fully quantum regime, blue lines), medium rate of dephasing (ENAQT regime, red lines) and high dephasing rate (classcial regime, green lines). High dephasing rate corresponds to $\gamma_{deph}=1000 s^{-1}$, and medium value correponds to the dephasing rates at which maximum particle current is obtained for the same value of disorder ($\gamma_{deph}=30,55,98 s^{-1}$, see Fig.~ \ref{fig:Disordered_ENAQT}).

 By comparing the quantum region (blue lines) of Fig.~~ \ref{fig:ENAQT_Mechanism} a,b and c, one can notice the onset of localization, as the population goes from being determined by the wave-function structure to having a disorder-driven gradient, which is due to the localization of the wave-functions. Surprisingly, the ENAQT and classical regimes are hardly affected by disorder, because dephasing leads to mixing of eigen-states, and hence the single eigen-value delocalization plays no role.


Fig.~ \ref{fig:ENAQT_Mechanism}d shows the particle current (blue line) of an ordered system (solid line) and disordered system (dashed line) as a function of dephasing. Red line shows  $\Delta_n=\frac{1}{L}\sum_i \left(n_i -\bar{n}\right)^2$, which describes the population spread along the chain\cite{zerah2018}. It is clear from Fig.~ \ref{fig:ENAQT_Mechanism}d that a minimal $\Delta_{n}$ is obtained at the same dephasing rate that maximum current is obtained, thus reflecting the population uniformiuzation mechanism, even at high disorder.

\prlsec{Interactions}\label{sec:Interactions} We now turn to discuss the interplay between interactions and dephasing. We consider an ordered system for simplicity (disorder does not changes the results qualitatively). To take exciton interactions into account, we add to the Hamiltonian an interaction term of the form $H_{int}=\sum_i U \hat{n}_i\hat{n}_{i+1}+h.c.$, where $ \hat{n}_i =e^\dagger_i e_i  $ and $U$ is the interaction strength (the interaction is limited to nearest-neighbors for simplicity). \\ \\ 

\begin{figure}[H]
    \centering
    \includegraphics[width=8truecm]{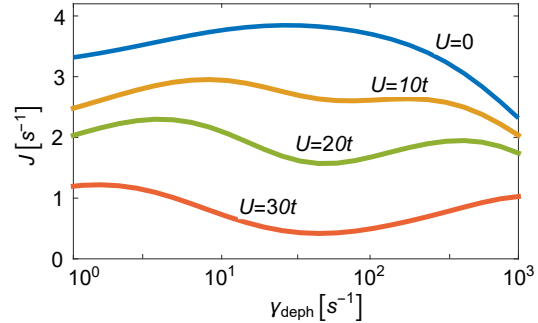}
    \caption{Particle current as a function of $\gamma_{deph}$ for different values of exciton interaction strength U=0,10t,20t,30t. A qualitative change of behaviour is observed, from ENAQT to environment-hampered transport. Other parameters are as in figure 2. }
    \label{fig:interactions}
\end{figure}

In Fig.~\ref{fig:interactions} we show the particle current as a function of dephasing rate, for different values of interaction strength $U=0,10t,20t$ and $30t$. The parameters were carefully chosen to be within the two-exciton limit (see appendix A and B).

In the absence of interactions (blue curve in Fig.~ \ref{fig:interactions}), ENAQT is observed even when two excitons are occupying the system. However, as the interaction strength increases (red, green and orange lines), the behavior changes qualitatively, reversing the trend from an environment-{\sl assisted} transport, to an environment-{\sl hampered} transport.

The origin of the interaction-induced reduction of current can be understood by first noting that it appears only when the rate of particle injection is large enough so that the particle current is dominated by two-exciton states. The addition of nearest-neighbor repulsive interaction induces an effective barrier that an incoming particle has to cross in order to reach the extraction site. One would thus think that it would have the same effect as disorder (i.e. only quantitative), but this is not the case. The reason is that the interaction-induced barrier is not random. Rather, it is structured in  such a way that the extraction site is always disfavored (see appendix F). Thus, dephasing fails to uniformize the population, and only reduces the extraction site population (and hence the particle current). 



\prlsec{Conclusion} In this letter, we have discussed the interplay between disorder, interactions and transport in open quantum systems. We showed that under certain conditions disorder can assist transport, and demonstrated that ENAQT persists even when the underlying hamiltonian system is deep in the disorder-induced localized regime. This implies that the mechanism for ENAQT is not directly related to localization, but rather comes from the competition between the tendency of the dephasing environment to uniformize of the particle density and the formation of a density gradient \cite{zerah2018}. The appearance of repulsive interactions changes the response of the system to a dephasing environment in a qualitative way, leading to an environment-hampered transport regime. Our conclusion can be readily tested with current experimental capabilities, whether the quantum particles are photons in a photonic lattice \cite{viciani2016, caruso2016} or qubits in ion traps \cite{maier2019}. Future studies will clarify the roles of long-range interactions and colored noise in contributing to ENAQT.

\appendix
 

 \section {Equation for exciton number}
 We start by evaluating analytically the average of the total exciton density. The basic Hamiltonian and formalism are described in the text, and we limit ourselves (for the sake of obtaining analytic results) to the single-exciton Fock space (i.e. we allow the exciton network to be either empty or occupied by a single Exciton). This approximation is, of course, invalid for a system with large exciton number, but allows for analytic analysis.

 The average number of excitons in a wire that is attached to source and drain by its edges is given by
 \begin{equation}\label{eq:s1}
 \braket{\hat{N}}=\mathrm{trace}(\rho \hat{n})=\sum_{i=1}^{L}\rho_{i,i}=1-\rho_0
 \end{equation}
 where $\hat{n}$ is the number operator, $\rho$ is the density matrix, $\rho_{i,i}$ is the diagonal element of site $i$, and $L$ is the number of sites.\\
 An expression for $\rho_0$ was derived by the relation \cite{zerah2018}
 \begin{equation}
 \frac{\gamma_{inj}}{\gamma_{ext}}=\frac{\rho_{ext}}{\rho_0}
 \end{equation}
 where $\gamma_{inj},\gamma_{Ext}$ are the injection and extraction rates, respectively. $\rho_0,\rho_{ext}$ are the empty state density, and the extraction site density.\\
 We use $\rho_{ext}$, derived in a previous publication \cite{zerah2018} and obtain
 \begin{widetext}
 \begin{equation}
 \rho_0= \frac{4t^2 \gamma_{ext}}{4t^2 \gamma_{ext}+ 4L\cdot\gamma_{inj}t^2+(L-1)\gamma_{ext}^2\gamma_{inj}+(L^2-L)\gamma_{inj}\gamma_{ext}\gamma_{deph}}~~,
 \end{equation}
 where $L$ is the number of sites in the wire.\\
 Replacing this expression with equation \ref{eq:s1}, we get (after some algebra) 
 \begin{equation}
 \rho_0=\frac{1}{1+L\cdot\frac{\gamma_{inj}}{\gamma_{ext}}+\frac{L-1}{4}\cdot\frac{\gamma_{ext}\cdot\gamma_{inj}}{t^2}+\frac{L^2-L}{4}\cdot\frac{\gamma_{inj}\cdot\gamma_{deph}}{t^2}}~~.
 \end{equation} \end{widetext}
 Substituting into equation \ref{eq:s1} leads to:\begin{equation}\label{eq:num_o_exci} \braket{N}= \eta\left(\eta+K^{-1} \right)^{-1} 
 \end{equation}
 
 where $\eta=\frac{\gamma_{inj}}{\gamma_{ext}}$ and
 \begin{equation} K=L+\frac{L+1}{4}\frac{\gamma_{ext}^2}{t^2}+\frac{L(L-1)}{4}\frac{\gamma_{ext}\gamma_{deph}}{t^2}~~.
 \end{equation}
 
 This result shows  that the most important parameter controlling the average exciton number is the injection-extraction rates ratio, $\eta=\frac{\gamma_{inj}}{\gamma_{ext}}$. If $\eta<<1 $ (as is the case in, e.g., natural exction transfer complexes in physiological conditions), then the system is occupied by very few excitons, and hence the single-exciton approximation is valid.

 The parameters of the system in the main text are chosen such that $\eta=1$, to mimic experimental conditions. In this case 
 \begin{equation}\label{eq:n_exc}
 \braket{N}=\frac{K}{K+1}~~,
 \end{equation} 
 and since $K$ is not small (it is always larger than $L$), the exciton density is close to $1$ (within the one-exciton approximation!) implying that the full Fock space is required. Nontheless, as we show in the next section, for non-interacting excitons, the difference between the high-density and low-density cases are only qualitative, and the use of the single-exciton approximation still points to the correct physical mechanism of ENAQT. 
 
 \section{Multiple excitons and disorder}
 Here we repeat the calculations shown in Fig.~2 and Fig.~3 of the main text, in a system which is limited to a very low number of excitons. All numerical parameters are the same, except the injection rate, which is taken to be $\gamma_{inj}=0.17s^{-2}$, namely a hundred times smaller. This gives $\eta=0.01$ and $\braket{n}\sim 0.1$, making the single-exciton approximation perfectly reasonable. 
 
 As can be seen form Fig.~\ref{fig3_2ex} the central features of the non-interacting system are still observed; both the non-monotonic dependence of current on disorder strength at low dephasing rate (Fig.~\ref{fig3_2ex} top panel) and the ENAQT behaviour and its persistence for high disorder (Fig.~\ref{fig3_2ex} bottom panel).

 \begin{widetext}
 \begin{center}
 	\begin{figure}[H]
 		\centering
 		\includegraphics[width=1\textwidth]{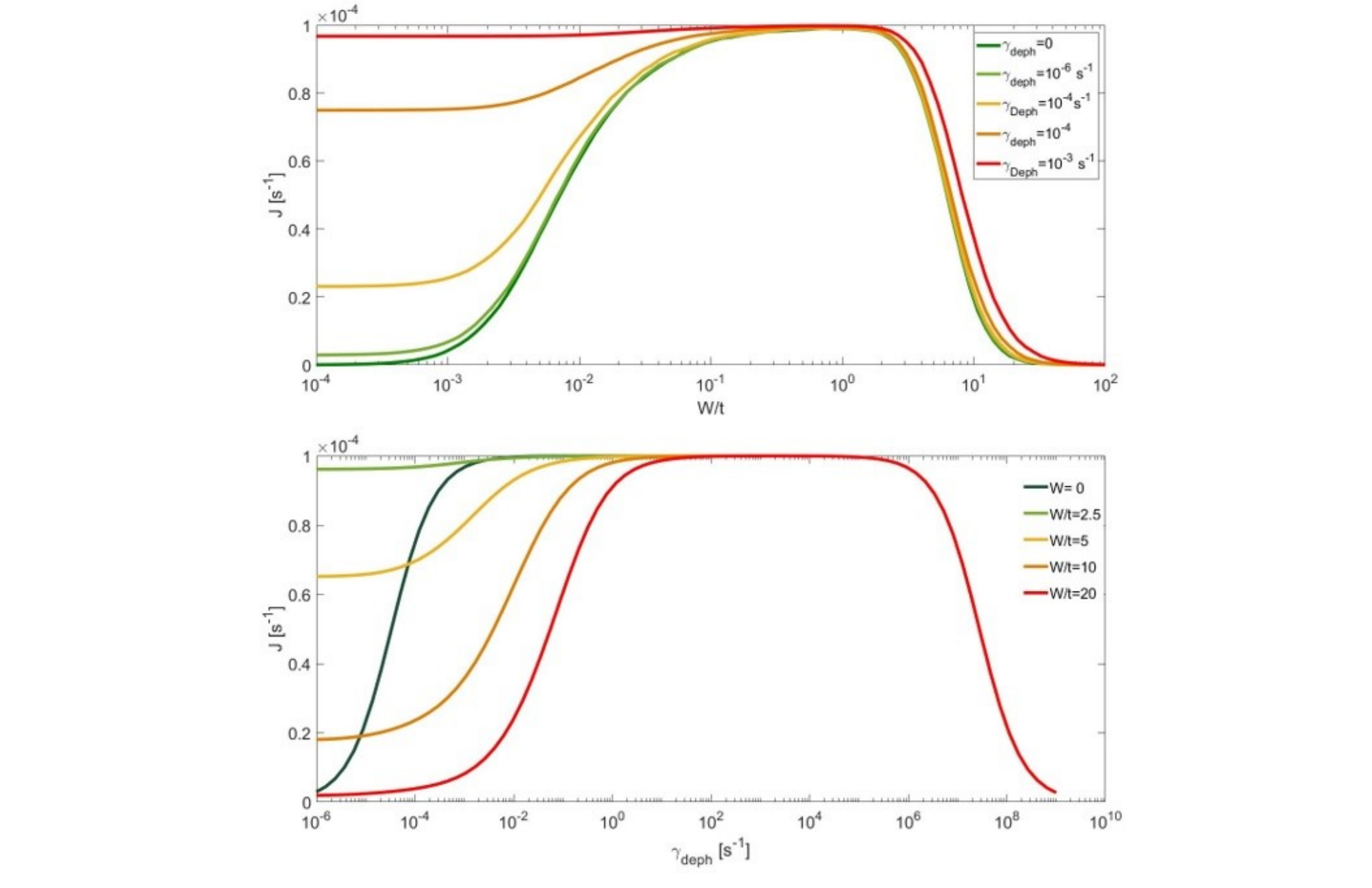}
 		\caption{Top panel: Current as a function of disorder strength for different dephasing rates (see legend) for injection to extraction rates ratio $\eta=0.01$ (in the low exciton population regime), showing similar behaviour to a system with higher exciton density, thus supporting the use of the single-exciton approximation in the mian text. Bottom panel: Current as a function of dephasing rate for differentdisorder strengths, showing again similar behavior as in the main text.  }
 		\label{fig3_2ex}
 \end{figure}\end{center}
 \end{widetext}
 
 \section{role of Disorder: detailed mechanism}
 Here we discuss in detail the mechanism for the enhancement of current by (static) disorder. 
 
\begin{center}
 	\begin{figure}[H]
 		\centering
 		\includegraphics[width=0.5\textwidth]{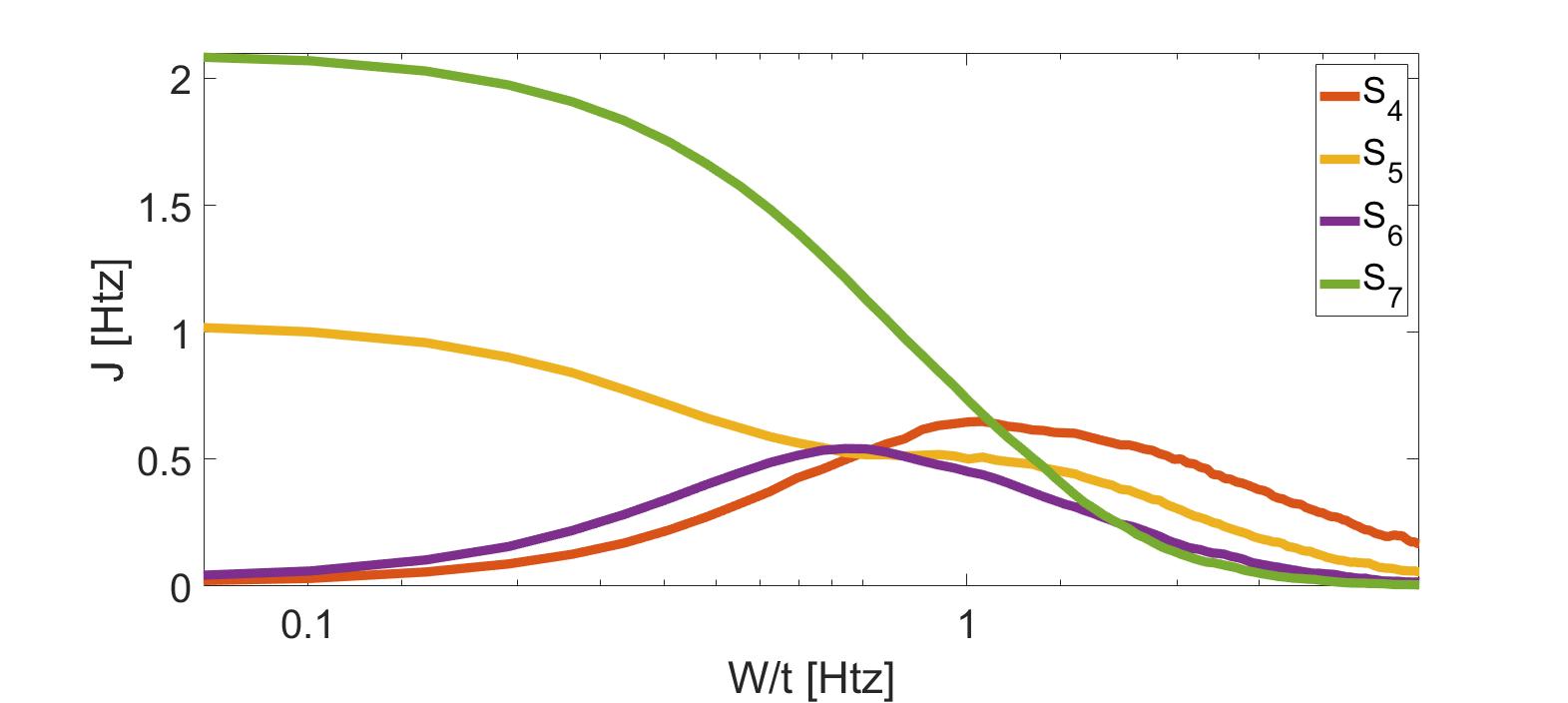}
 		\caption{Particle current as a function of disorder (W/t), for a chain of seven sites. Excitation takes place at the first site, different lines depict different extraction sites: 4-6 (red, orange, purple and green).}
 		\label{fig_difExt}
 \end{figure}\end{center}

 Figure \ref{fig_difExt} shows the current as a function of disorder ($W$) for a chain of seven sites (same numerical parameters as in the main text), where each line shows the current for different extraction site: $s_i=4,5,6,7$ (red to green lines). As can be seen, maximum current is obtained for systems with extraction sites positioned at sites $4$ and $6$, while systems that their extraction sites are in sites $7$ and $5$ exhibit monotonic behaviour with increasing disorder.
 
 Figure \ref{Figs2} shows the particle density in the system in real space (a,b) and in eigen-space (c,d) for a system coupled to the drain by site 5 (b,d) and site 6 (a,c). Each line depicts the particle density for different disorder strength, $W/t=0,1,5,20$ (color code in the figures) at zero dephasing, $\gamma_{deph}=0$.

 Since the current of a system is proportional to the density at the extraction site \cite{zerah2018}, it is again useful to see how the populations behave under increasing disorder. Fig.~\ref{Figs2}(a-b) shows the particle density as a function of position along the chain for chains with extraction sites at $i_{etx}=5$(a) and $i_{etx}=6$(b), for different values of disorder $W/t=0,1,5,20$. In both cases, it seems that there is a disorder-induced crossover from a population which is distributed along the chain at weak disorder (blue lines) to a density which is localized at the injection site and decays at the extraction site (green lines) 
 
 So why is there an inhomogeneous disorder-dependence of current in the configuration of Fig.~\ref{Figs2}(b) and not in Fig.~\ref{Figs2}(a)? a hint comes from looking at the population of {\sl eigen-states} of the different systems, which is shown in Fig.~\ref{Figs2}(c-d) for extraction sites 5 and 6 respectively. As seen, for $i_{ext}=6$ the population in eigen-space is localized at one eigen-state for the clean system (Fig.~\ref{Figs2}d, blue line), but for $i_{ext}=5$ the clean system is spread over many eigenstates (Fig.~\ref{Figs2}c, blue line). For both cases, as disorder increases the system becomes delocalized in eigen-space. 
 \begin{widetext}
 \begin{center}
 	
 	\begin{figure}[H]
 		\centering
 		\includegraphics[width=1\textwidth,scale=0.9]{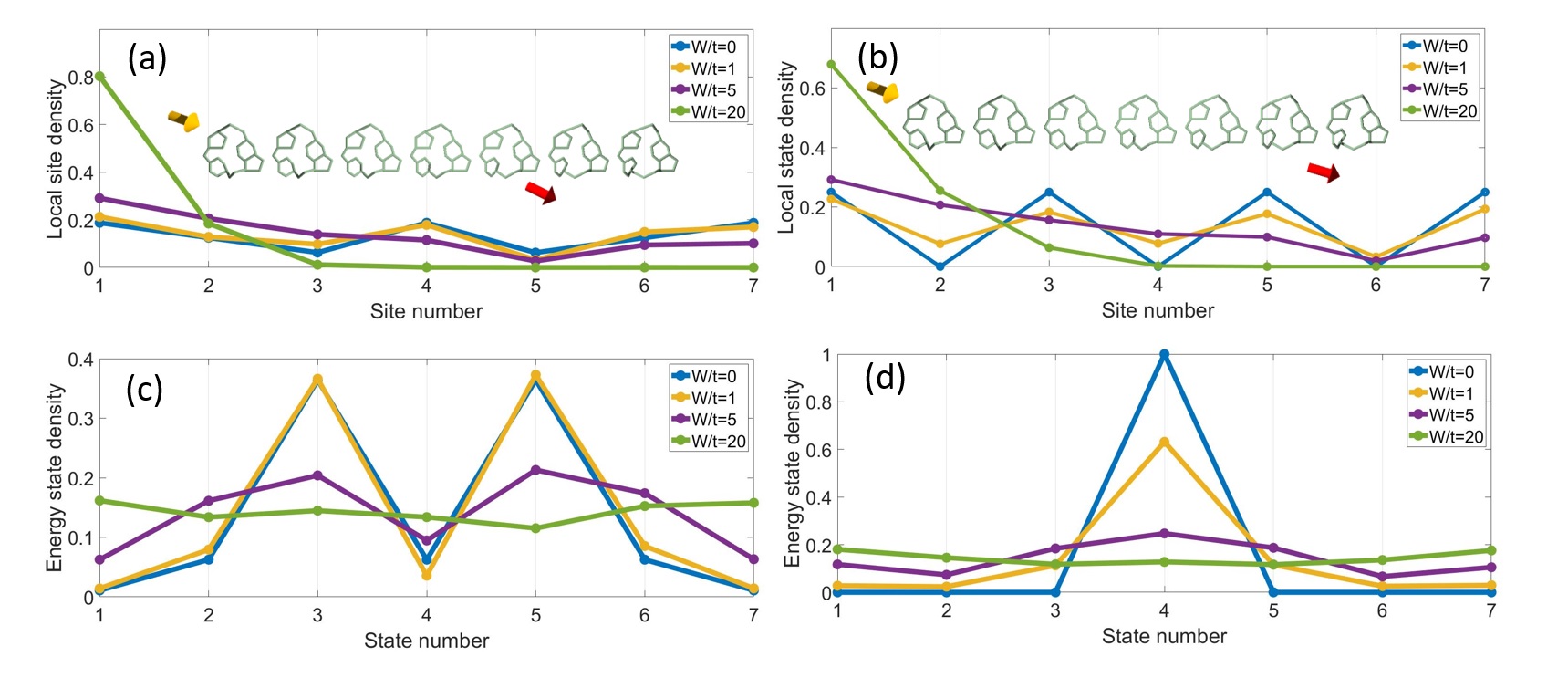}
 		\caption{Local space (a,b) and energy space (c,d) of seven sites wire with extraction from site 6 (b,d) and extraction from site 5 (a,c). Different color lines resembles different disorder strength}
 		\label{Figs2}
 \end{figure}\end{center}
  \end{widetext}
 
\begin{center}
 	
 	\begin{figure}[H]
 		\centering
 		\includegraphics[width=0.5\textwidth]{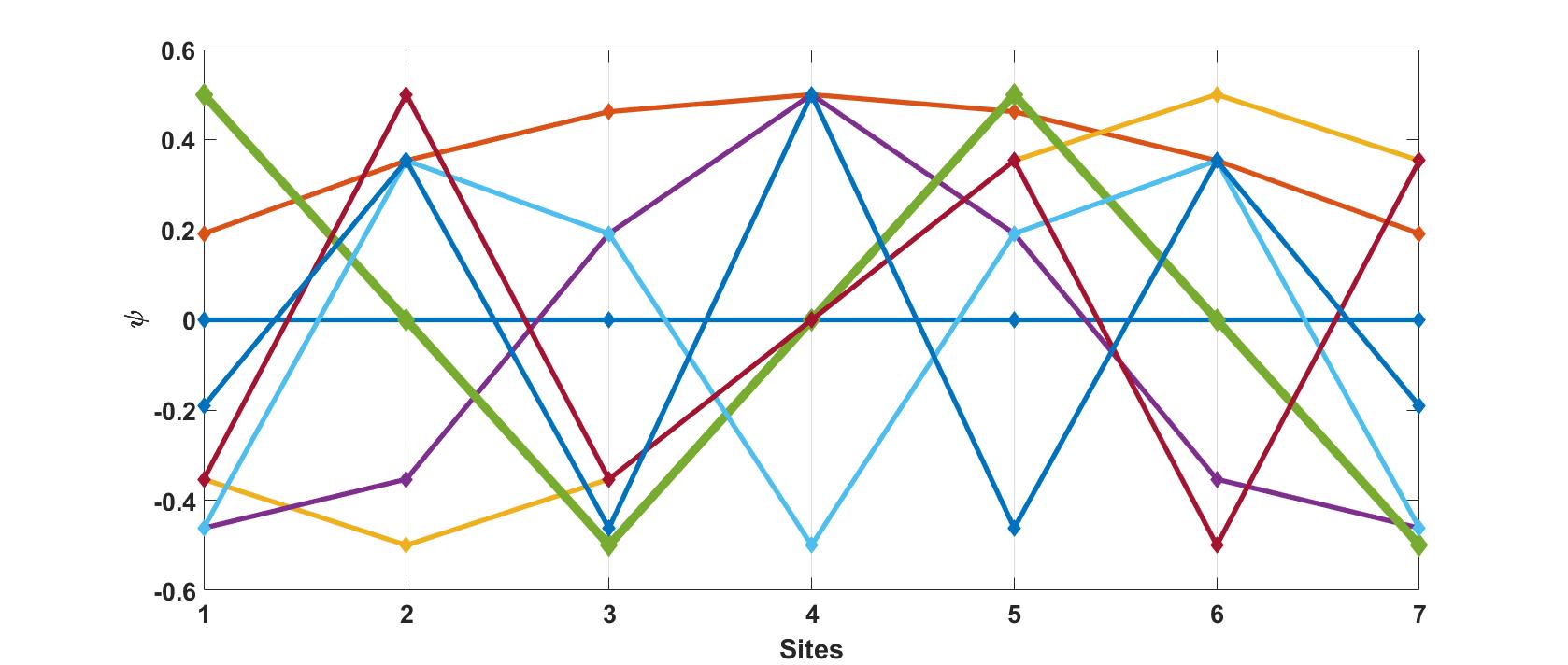}
 		\caption{Eigen-functions of seven sites chain }
 		\label{FigsEigen}
 \end{figure}\end{center}

 How is this related to inhomogeneity in the current vs disorder? to see this, we note that when there is inhomogeneity in the current vs disorder, the clean system is localized in eigen-space, and the specific eigen-state chosen is such that it has a node at the extraction site. This can be seen in Fig.~\ref{FigsEigen}, where we plot the eigen-states of the 7-site system. One can see that sites 4 and 6 have nodes, and these are the sites where the disorder-enhanced current appears (Fig.~\ref{figs1}). 
 
 When there is no node at the extraction site, the system occupies eigen-states in such a way as to reduce population at the extraction site, by choosing a combination of eigen-states. Then, for each of the eigen-states, the current is reduced upon increasing disorder, due to the real-space localization of the wave-functions. 
 
 \section{Disorder in a chain with long-range coupling}
 
 In Ref.~\cite{maier2019}, the chain is characterized by long-range hopping matrix elements of the form $t_{ij}=\frac{t_0}{|i-j|}$. Here we show that essentially all our conclusions, which were drawn for nearest-neighbor couplings, equally apply to the long-range-hopping chains. We evaluate a chain with similar numerical parameters as in the text (and extraction site at site 6), but take the long-range hopping term above. 
 
\begin{center}\begin{figure}[H]
 		\centering
 		\includegraphics[width=0.5\textwidth]{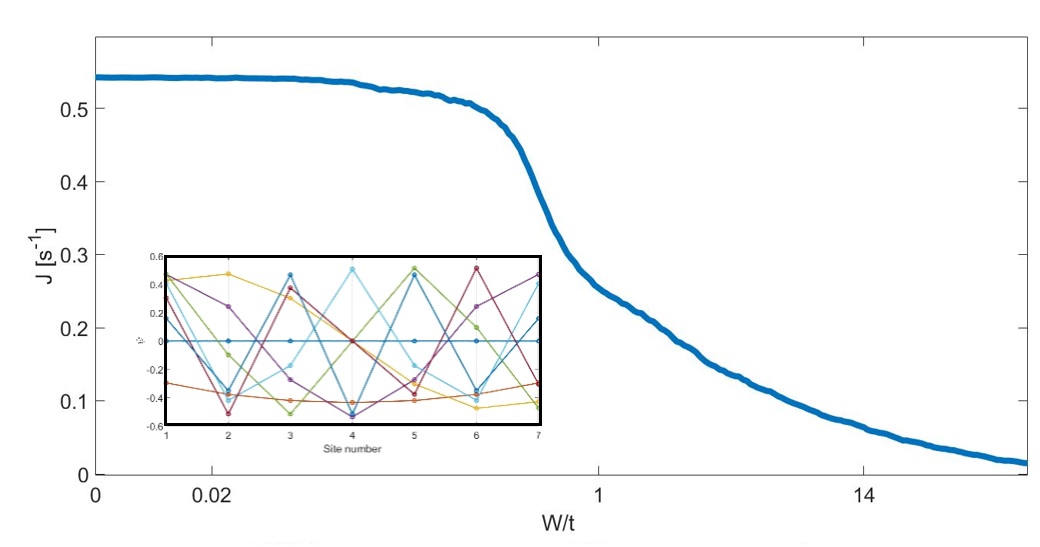}
 		\caption{Particle current as a function of disorder for a chain of seven sites with long range couplings. Inset: eigen-states of the Hamiltonian with long-range couplings.  }
 		\label{fig:longInt}
 \end{figure}\end{center}

 We start by evaluating the current as a function of disorder without dephasing, as plotted in Fig.~\ref{fig:longInt}. There is no disorder-enhanced current, and the reason is that due to the long-range nature of the hopping, there is no longer a node in the eigen-states at the extraction site, as can bee seen in the inset. This is fully consistent with the discussion in the main text.

\begin{center}\begin{figure}[H]
 		\centering
 		\includegraphics[width=0.5\textwidth]{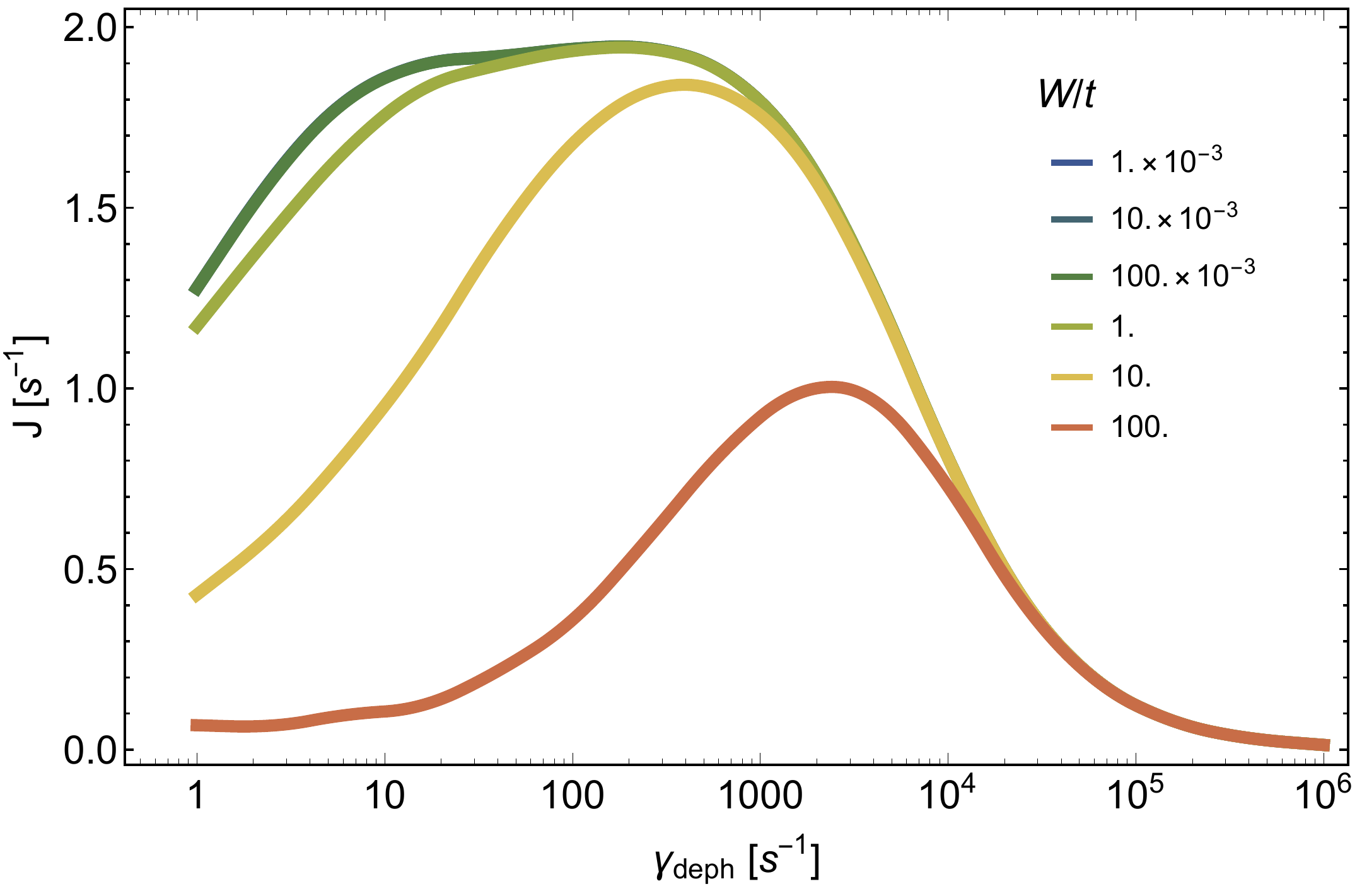}
 		\caption{Current as a function of dephasinf rate for different values of disorder, for a chain with long range couplings. }
 		\label{fig:SM_Jvsgamma_longrange}
 \end{figure}\end{center}

\begin{center}\begin{figure}[H]
 		\centering
 		\includegraphics[width=0.5\textwidth]{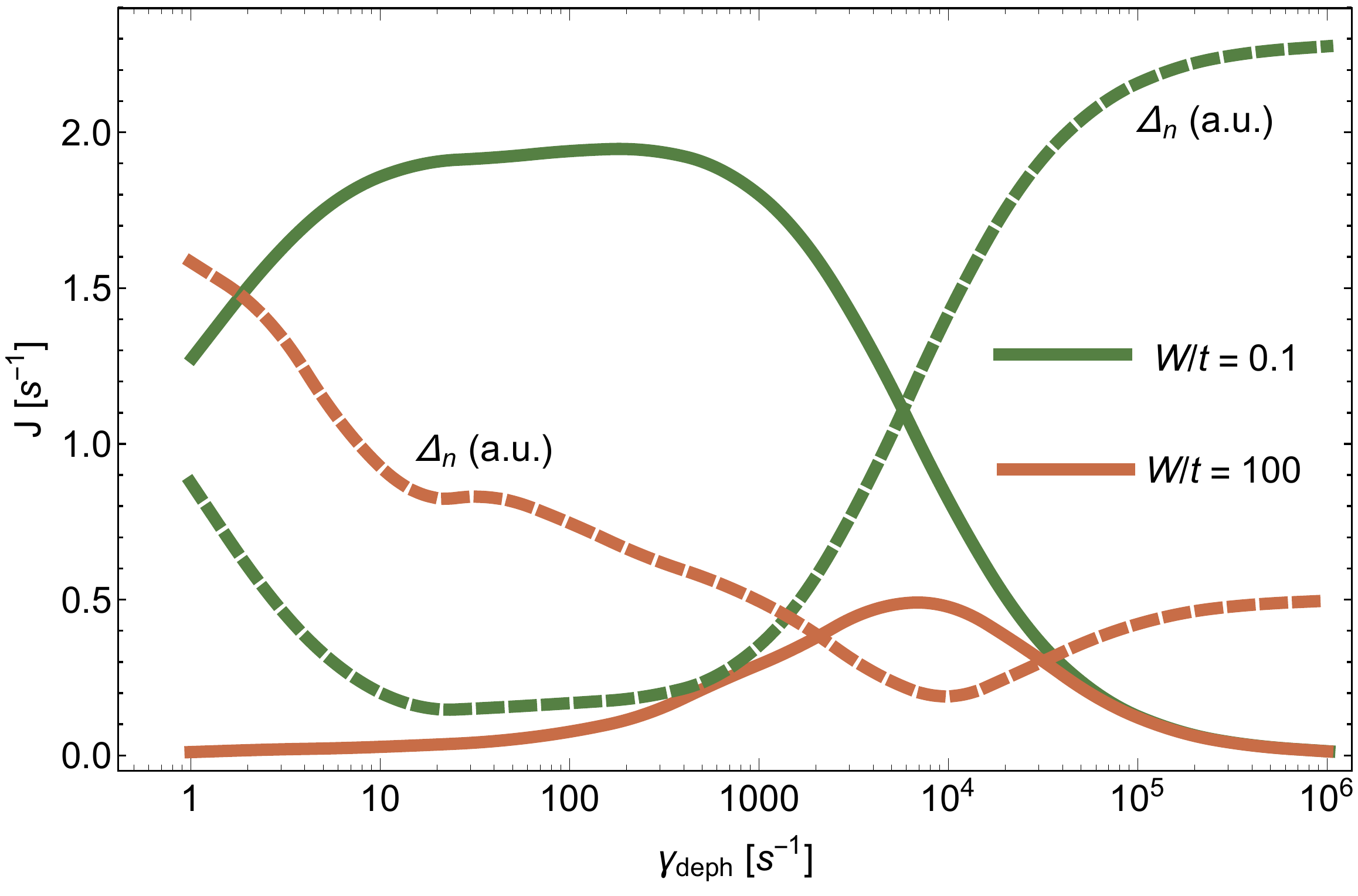}
 		\caption{Solid lines: current as a function of dephasing for different values of disorder (same data as in Fig.~\ref{fig:SM_Jvsgamma_longrange}). Dashed lines: the population spread parameter, $\Delta_n$, showing a minimum value when the current is maximal, thus exhibiting the same origin of ENAQT for chains with long-range couplings.    }
 		\label{fig:SM_Jvsgamma_longrange_II}
 \end{figure}\end{center}

 In Fig.~\ref{fig:SM_Jvsgamma_longrange} we show current as a function of dephasing rate for different values of disorder strength (legended in the figure), same as Fig.~3 in the main text. As seen, the long-range system exhibits the same phenomenology, Showing ENAQT even for strong disorder. In Fig.~\ref{fig:SM_Jvsgamma_longrange_II} the current vs dephasing and the population spread parameter, $\Delta_n$ (in arbitrary units, as in Fig.~4d of main text), are plotted for two values of disorder strength, showing that also here ENAQT is related to minimization of the population spread. Simply put, the population uniformization mechanism works also for the case of long-range hopping, although the structure of the eigen-states is different. This is in line with the conclusions of Ref.~\cite{zerah2018}.

 \section{ENAQT in an interacting system}
 In order to understand how interactions lead from ENAQT to dephasing-induced current reduction, we start by looking at the total Hamiltonian of the interacting system, limited to Fock space of $\leq 2$ particles.  The tight-binding Hamiltonian (shown in equation 1 in the main text)  includes the presence of two excitons in the wire (orange block), the presence of one exciton (gray block) and the possibility of no excitons (top left element). The interaction between different blocks (Fock subspaces with different particle numbers) is only possible via excitation or annihilation process, induced by the source and drain Lindblad operators (and not by coupling element in the Hamiltonian). 
 
 In the singly-occupied sub-system (SOSS), ENAQT takes part at an intermediate rate of dephasing through the population uniformization mechanism \cite{zerah2018} ).  The doubly-occupied sub-system (DOSS) is more complicated than the SOSS. Looking only at the DOSS Hamniltonian, one can see three elements which differ from the SOSS:  (1) The quasi structure - simply put, if this were a single-particle Hamiltonian, it would have a different geometry than the SOSS,(2) Multiple injection and extraction sites and (3) Non-uniformity in the on-site energies (due to interactions). Accordingly, despite their similar mechanism (and physical origin), SOSS and DOSS operate differently in a way that give rise to a new effect that responsible of the current reduction. \\
 
\begin{figure}[H]
 	\centering
 	\includegraphics[width=0.5\textwidth]{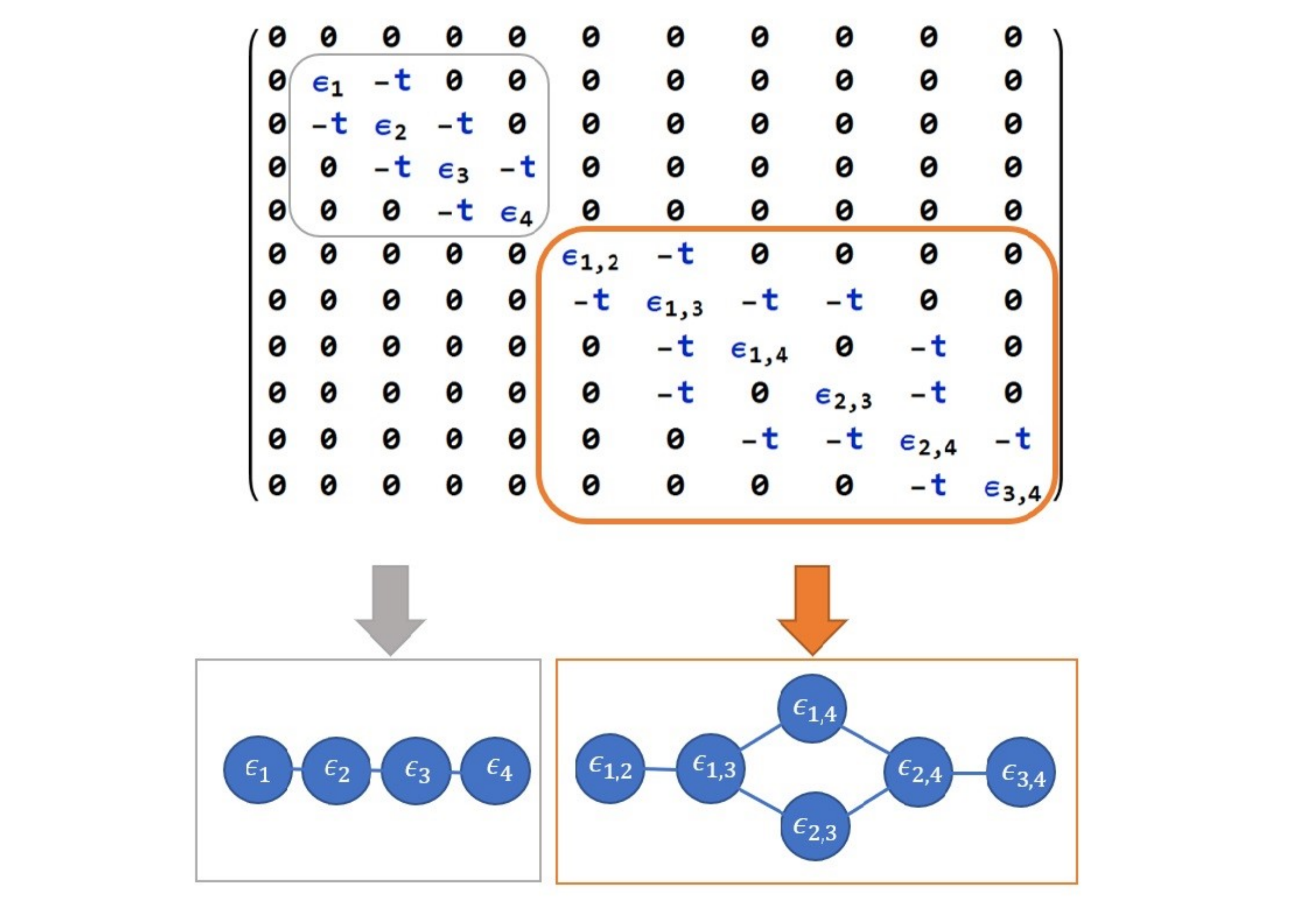}
 	\caption{An example of quasi-structure due to interactions: The Hamiltonian for 4-site chain (restricted to occupation of up to two particles) is shown on the top of the figure. It contains two blocks: the SOSS (top gray rectangle), resembling a 4 sites wire Hamiltonian (bottom gray rectangle), and the DOSS (top orange rectangle), which can be mapped to an effective single-particle system (bottom orange rectangle).}
 	\label{figs1}\end{figure}

 We first note that one can understand the origin of the dephasing-induced current reduction by noting that, at least form the mathematical viewpoint, the contribution of the SOSS and DOSS to the current is similar, and one can think of the DOSS as a complicated single-particle system which has the DOSS Hamiltonian. Then, one needs to understand how the three elements described above affect the current in the presence of dephasing. 
 We start with the effective network morphology of the DOSS. Fig.~\ref{figs1}
 shows, for example, the  Hamiltonian matrix of four-sites-wire. Orange rectangle emphasizes the DOSS part, and the orange arrow points on the quasi structure that is formed from the locations of the coupling elements in the Hamiltonian matrix. Since the site populations have a strong dependence on the morphology of the network, which, in turn affect the particle current, this effective structure transfers particles significantly different then the simple wire depicted by the SOSS (shown in gray rectangle in figure \ref{figs1}).
 
 The second contribution is the multiple injection and extraction sites. In the SOSS, these are the sites that are locally attached to the source and drain, while in the DOSS, each state that includes the injection (extraction) site is coupled to the source (drain), i.e for each injection (extraction) site in the SOSS, there are $L-1$ injection (extraction) sites in the DOSS. As an example, site $1$ in the DOSS shown in figure \ref{figs1} is the injection site, accordingly, in the DOSS, injection states are $\rho_{1,2},\rho_{1,3},\rho_{1,4}$.  The appearance of many injection and extraction sites raises the complexity in the classic limit, where the occupation of sites is organized to form a linear density gradient (Fick's law \cite{zerah2018}).  In case of more then one injection site (or extraction site) more linear gradients are formed, leading to a non-intuitive total occupation organization.
 
\begin{figure}[H]
 	\centering
 	\includegraphics[width=0.5\textwidth]{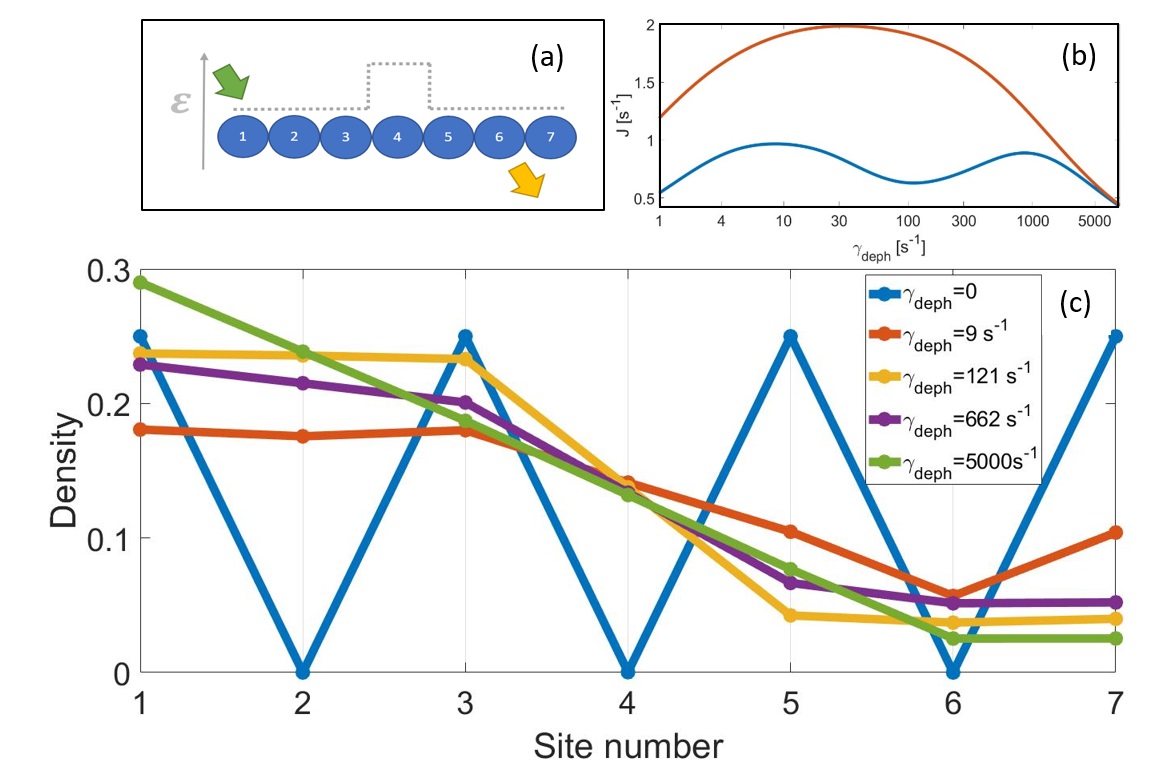}
 	\caption{The effect of an energy-barrier: (a) schematic description of the system-Single occupied seven sites chain with energy barrier  at the fourth site. Excitation takes place at the first site and extraction at sixth site, as shown by green arrow and yellow arrows, respectively. (b) Blue curve shows the particle current of the system in (a), red curve shows the particle current of a uniform seven sites chain.(c)
 		Site densities, shown for the quantum limit (blue curve), intermediate domains (red and purple curves), minimum current point 
 		(orange curve) and the classic limit (green curve). The ENAQT regime is affected from the energy barrier, where the density diminishes for the sites that are positioned after the bump (site 4). Further increase in the  dephasing rate (the classical limit) results in the formation of a linear gradient (green curve) between the injection and extraction sites, while it path through intermediate configuration (purple curve). (Numerical parameters: $\gamma_{inj}=17 s^{-1},\gamma_{ext}=17ps^{-1},U=50t,\epsilon=300t,t=144 s^{-1}$ )   }
 	\label{figs2}\end{figure}

 The final (and most significant) contribution is the non-uniformity of the on-site energies that is created as a result of the interaction between particles, which - in the mapping to a SOSS - is interpreted as an extra energy in the relevant double occupied states (those who include neighboring sites). Consequently, an energy barrier is created. In order to study this effect, we analyze a single-particle system which has a similar energy landscape, i.e. a chain of seven sites which has one "energy barrier" in site 4, as demonstrated schematically in figure \ref{figs2}a, to mimic the inherent energy barrier which is a result of the interactions. While the full interacting system has multiple energy barriers, considering one is enough to provide intuition into the origin of the dephasing-induced current reduction.

 In Fig. \ref{figs2}b the particle current as a function of dephasing for the energy barrier chain (blue line), and compared to the uniform chain (red line). The density along the chain (with energy barrier at site 4) is plotted in Fig.~ \ref{figs2}c.
 
 The picture that comes out of these data is the following. When an energy barrier is present, in the absence of dephasing, the barrier site is under-populated (compared to the same chain with no barrier). As dephasing is increased, due to the position of the source and drain, the situation becomes different than the regular popularization uniformization; as the barrier site population increased, the population of sites between the barrier and the extraction site decreases. In other words, the barrier "disconnects" the sites close to the drain from the rest of the chain, and as dephasing induces eigen-state delocalization and mixing of level, the effect on states close to the drain is actually to "feel" the drain even more, and have their population reduced. 
 
\begin{figure}
 	\centering
 	\includegraphics[width=0.5\textwidth]{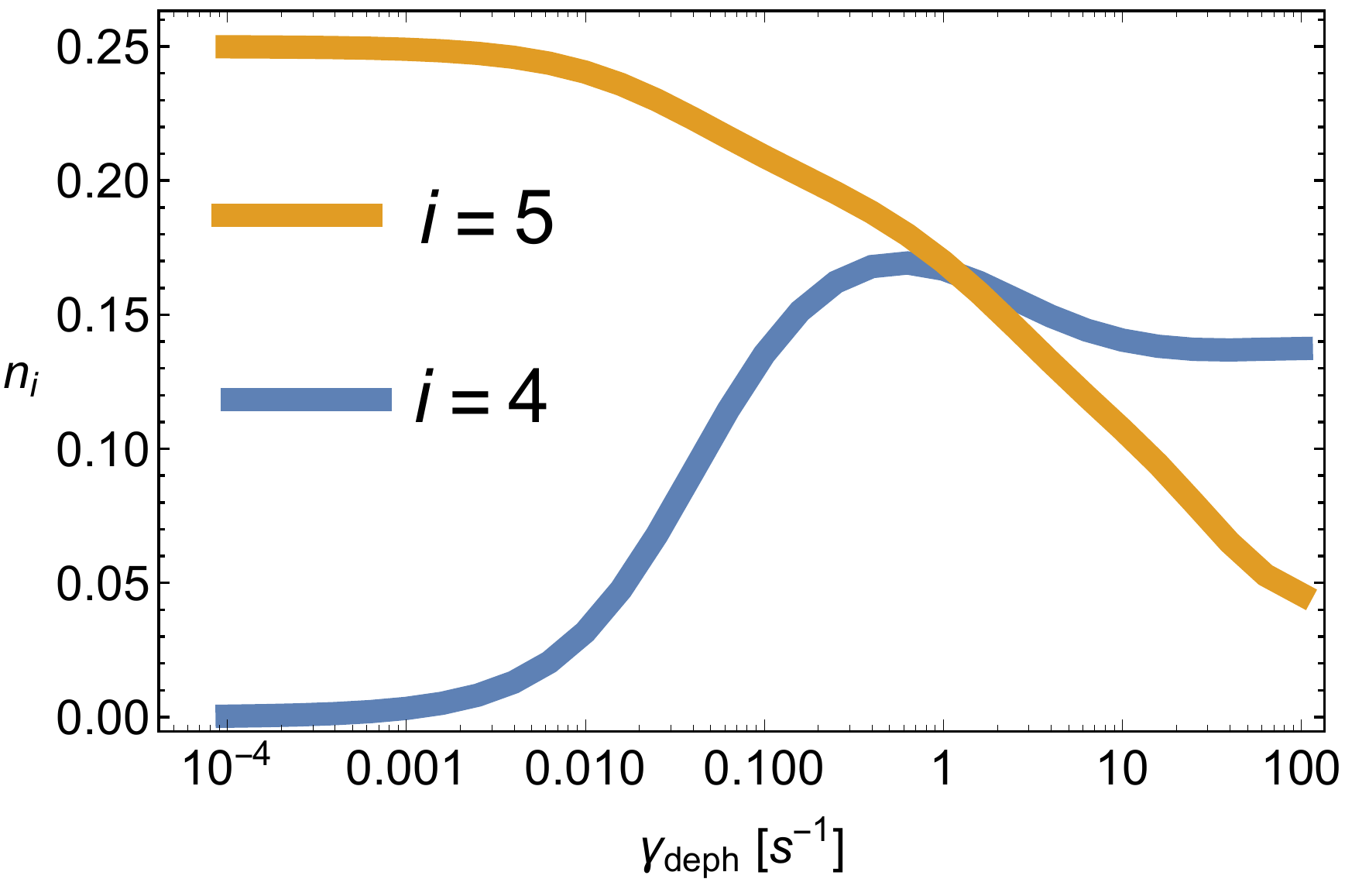}
 	\caption{Population of energy barrier site (site 4, blue curve) and a site that is positioned between the barrier and the drain (site 5, orange curve), is shown as a function of dephasing}
 	\label{S7density}\end{figure}

 An example of this can be seen in Fig.~\ref{S7density}, the population of site 4, where the energy barrier sits (blue line) and site 5 (yellow line) are plotted as a function of dephasing rate. As seen, the population of the barrier site increases with dephasing, as the population of site 5, which is between the barrier and the drain, decreases. This is typically the case (although, for the non-interacting chains, geometry also plays a role).
 
\begin{figure}
 	\centering
 	\includegraphics[width=0.5\textwidth]{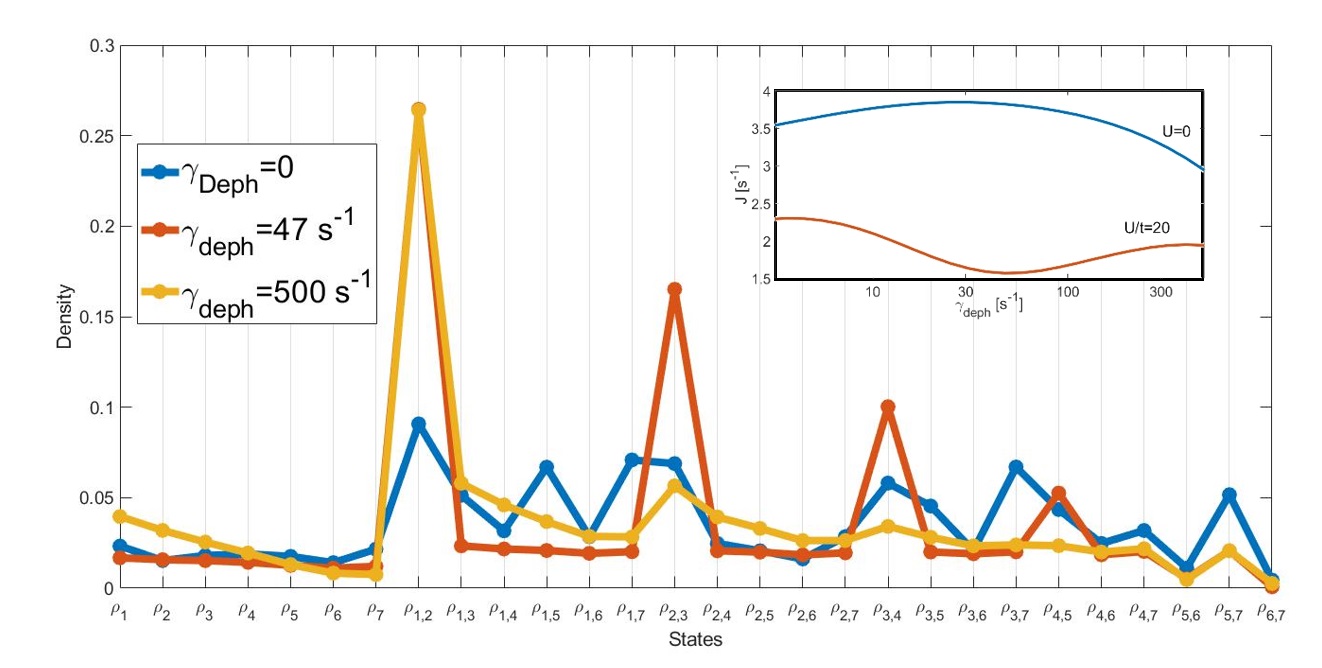}
 	\caption{Diagonal elements of the density matrix (i.e. population of different states) for different dephasing rates ($\gamma_{deph}=0,47$ and $500$ s$^{-1}$, represented by blue, red and yellow curves respectively). States with two indices belong to the doubly occupied Fock sub-space. The occupation of states  $\rho_{1,2},\rho_{2,3}$  due to the presence of an effective energy barrier. In the classical limit (blue curve) the linear gradient is more dominant. Accordingly, the particle current decreases significantly in the ENAQT regime, as shown in the inset (red curve) . }
 	\label{figs3}\end{figure}

 The summation of all the contributions discussed above comes into play in generating the environment-hampered transport in the presence of interactions. An example can be seen in Fig.~\ref{figs3}, where we plot the values of the different density-matrix elements for three values of disorder. The elements with two indices correspond to the "population" of states with two particles. The states with adjacent indices correspond to Hamiltonian states with an "energy barrier" due to the interaction. 
 

 Finally, we note that since the DOSS and the SOSS operate (and contribute to the current) as independent sub-systems, the final particle current is dictated by the combination of these two sub-systems. Their contribution to the total current will depend on whether or not the system has one or two particles in it. This, in turn, is determined by the injection to extraction rates ratio (IERR),  $\frac{\gamma_{inj}}{\gamma_{ext}}$ . If the IERR is large, DOSS is more likely, hence more dominant and consequently, and would give rise to the reduction of current. On the other hand, for small IERR, only the SOSS is dominating the transport, and the current reduction turns into a marginal effect.

\begin{acknowledgments}
We thank Y. Meir, A. Vardi and S. Sarkar for valuable discussions. EZH Acknowledges support from the Ilse Katz center interdisciplinary fellowship. This work was supported in part by the ISF grant 1360/17.
\end{acknowledgments}

\bibliographystyle{apsrev4-1}

%

\end{document}